\documentclass[twocolumn,showpacs,preprintnumbers,amsmath,amssymb]{revtex4}

\usepackage{graphicx}
\usepackage{dcolumn}
\usepackage{bm}

\begin{document}



\title{An experimental route to spatiotemporal chaos in an extended 1D oscillators array}



\author{M.A. Miranda}
\email[e-mail address:\ ]{montse@fisica.unav.es}
\affiliation{Dept. of Physics and Applied Mathematics, Universidad de Navarra. Irunlarrea s/n, E-31080 Pamplona, Spain}

\author{J. Burguete}
\email[e-mail address:\ ]{javier@fisica.unav.es}
\affiliation{Dept. of Physics and Applied Mathematics, Universidad de Navarra. Irunlarrea s/n, E-31080 Pamplona, Spain}


\begin{abstract}
We report experimental evidence of the route to spatiotemporal chaos in a large 1D-array of hotspots in a thermoconvective system. Increasing the driving force, a stationary cellular pattern becomes unstable towards a mixed pattern of irregular clusters which consist of time-dependent localized patterns of variable spatiotemporal coherence. These irregular clusters coexist with the basic cellular pattern. The Fourier spectra corresponding to this synchronization transition reveals the weak coupling of a resonant triad. This pattern saturates with the formation of a unique domain of great spatiotemporal coherence. As we further increase the driving force, a supercritical bifurcation to a spatiotemporal beating regime takes place. The new pattern is characterized by the presence of two stationary clusters with a characteristic zig-zag geometry. The Fourier analysis reveals a stronger coupling and enables to find out that this beating phenomena is produced by the splitting of the fundamental spatiotemporal frequencies in a narrow band. Both secondary instabilities are phase-like synchronization transitions with global and absolute character. Far beyond this threshold, a new instability takes place when the system is not able to sustain the spatial frequency splitting, although the temporal beating remains inside these domains. These experimental results may support the understanding of other systems in nature undergoing similar clustering processes. 
\end{abstract}


\maketitle


\section{INTRODUCTION}
\label{intro}

Understanding complex systems in physics, chemistry, biology and neuroscience is a topic of theoretical and experimental relevance, particularly when we are dealing with the route to spatiotemporal chaos in spatially extended systems. In this route, the existence of various subsystems or domains with spatiotemporal synchronization, coherent patterns, allows to handle a description in terms of clusters. Accordingly, we find cluster dynamics in networks of coupled oscillators such as: reactive-diffusion systems in chemistry~\cite{Vanag00,Kiss02}, multicellular tissues in neurobiology~\cite{Jampa07,Rajesh07} and coupled arrays of oscillators in physics~\cite{Kozyreff00,Wiesenfeld96}.
Although the scientific community has kept a great attention to complex dynamics during the last decades, a universal route to chaos in extended systems has not been already described, not even understood, in contrast with the theoretical and experimental effort devoted to confined systems. Our aim is to characterize experimentally the dynamics of a 1D-spatially extended system in the weak turbulence regime where a spatiotemporal splitting is produced inside synchronized clusters. Theoretically, Hohenberg and Shraiman approached the understanding of spatiotemporal chaos with the generalized Kuramoto-Sivashinsky equation~\cite{Hohenberg89} and also there are several recent numerical works behind this purpose in the field of complex networks~\cite{He06,Gomez07,Marti04}.\par
The experimental results reported here are enclosed in the field of extended, dissipative and complex systems involving many degrees of freedom. Our experiment is a quasi-1D thermoconvective system, an array of nonlinear coupled oscillators consisting of ascending convective cells (hotspots) exhibiting a spatiotemporal beating regime of the type of a zig-zag pattern (ZZ). It is characterized by the presence of stationary clusters, developed from a previous mixed pattern of irregular clusters. If we consider that strong nonlinear couplings between amplitude and phase might be neglected, the 1D-array is displaying ``phase locking phenomena''~\cite{Pomeau86}. We try to approach a phase description to follow the dynamics far from the threshold of the primary bifurcation. A particular point of view is also tackled on the phase synchronization theory of N-coupled oscillators which was firstly introduced by Winfree~\cite{Winfree67,Winfree90} and later developed by Kuramoto~\cite{Kuramoto03}. Modelling extended 1D-system of N-coupled oscillators undergoing global phase instabilities under certain constraints, such as weak coupling between oscillators subjected to the same driving force, has allowed to build rich phase spaces~\cite{Golomb92}.\par

1D-experiments are particularly interesting for the diversity of patterns they exhibit, for example in: thermoconvection~\cite{Burguete03,Burguete93,Maza94,Alvarez97,Ringuet93,Ringuet99,Gouesbet00,Pastur03}, electroconvection~\cite{Pastur01,Joets89}, the printer's instability~\cite{Pan94,Giorgiutti95}, directional solidification~\cite{Simon88}, directional viscous fingering~\cite{Bellon98}, circular liquid column arrays~\cite{Brunet04,Brunet101,Counillon98}, the Taylor-Dean system~\cite{Bot98}. Theoretical approximations to 1D-oscillatory instabilities have already been successfully modelled for supercritical bifurcations with the Kuramoto-Shivashinsky phase equation (KS)~\cite{Misbah94}, and for subcritical bifurcations with a Ginzburg-Landau equation adding a quintic stabilizing term~\cite{Deissler94,Deissler98}. Other 1D-localized patterns such as solitons, and spatiotemporal intermittency, have been modelled using a modified Swift-Hohenberg equation~\cite{Gil99} and the KS~\cite{Brunet07}. In a previous work~\cite{Miranda08} we had characterized quantitatively primary and secondary instabilities for the same experimental cell. In this research we reported the existence of a mixed state for thin layers from a stationary cellular pattern (ST). This mixed pattern (ST/ALT) consists of localized oscillatory domains in an alternating pattern (ALT) coexisting with the basic cellular pattern, ST, which remains always stable. Depending on the values of the control parameter, the new unstable pattern shows a different pinning to the underlying stationary cellular pattern ST, and thus, there might be to two different kinds of fronts: fluctuating and stationary.\par

Mixed states in weak transitions to chaos have been usually identified as spatiotemporal intermittency regimes i.e. the Rayleigh-B\'enard convection in an annular gap~\cite{Daviaud92} and the Faraday experiment~\cite{Kudrolli96}. A similar behavior is found in experiments with open shear flows where the basic state is always stable, such as plane Couette~\cite{Bottin98}, circular Couette and Poiseuille~\cite{Coles}. These instabilities have in common a ``subcritical branch'' sent to infinity~\cite{Saarloos03}. In this sense, bistability in these systems is understood as the coexistence between ``patches'' displaying the new pattern and the original one.\par

Experimentally, certain extended systems in closed flows show a higher complexity stage before achieving spatiotemporal chaos or defect mediated turbulence: regimes with temporal or spatiotemporal frequency splitting. For example, the dynamics of a 1D-front in electroconvection with liquid crystals~\cite{Pastur01,Deissler94} shows time biperiodicity, and in the Taylor-Dean system~\cite{Bot98} spatiotemporal biperiodicity appears. 
In 2D systems, spatiotemporal chaos via a zig-zag instability has been experimentally observed in liquid crystals~\cite{Ribotta86,Plaut98,ThHenriot,Dennin98} and it has been successfully modelled in the frame of the Ginzburg-Landau equation for a weak nonlinear regime~\cite{Oprea07} and for a selective forcing~\cite{Henriot03}.\par

 According to the results reported here, we have no knowledge of any previous experimental work of a zig-zag regime (ZZ pattern) in a 1D-convective system caused by a spatiotemporal beating, and developed from an irregular clustering process. So, this report is the first experimental evidence in hydrodynamics of an array of oscillators displaying clusters, irregular and stationary, as the result of an increasing coupling interaction between hotspots.\par

We report a cascade of secondary bifurcations in a quasi-1D convective system, a rectangular layer of fluid opened to the atmosphere under a 1D-heating at the center of the bottom plate. For a fluid with Prandtl number ($\equiv$ {\sl viscous diffusivity/ thermal diffusivity}) $Pr \approx$ 75, the competing effects between buoyancy and capillarity account for a B\'enard-Marangoni convection mechanism.  For a fixed depth ($d =$ 3 mm), as we increase the control parameter (the vertical temperature difference $\Delta T_v$), the system undergoes a subcritical instability towards a mixed pattern of irregular clusters from a stationary cellular pattern. Further on, a spatiotemporal beating regime settles down, this consists of one or two large sized clusters with stationary fronts. The quantitative analysis by complex demodulation techniques shows a continuous evolution of the selected order parameter, the amplitude of the critical modes, from which the spatiotemporal beating regime is found to be supercritical. Both oscillatory instabilities to the mixed pattern (ST/ALT) and to the beating regime (ST/ZZ), have absolute and global nature. We will focus on the behavior of the critical modes nearby the thresholds of these transitions and describe the subsequent lost of spatial frequency splitting of the oscillatory modes (ST/DW) in a new transition.\par

\section{EXPERIMENTAL SETUP}
\label{sec:setup}

\begin{figure}
\includegraphics[width=8cm]{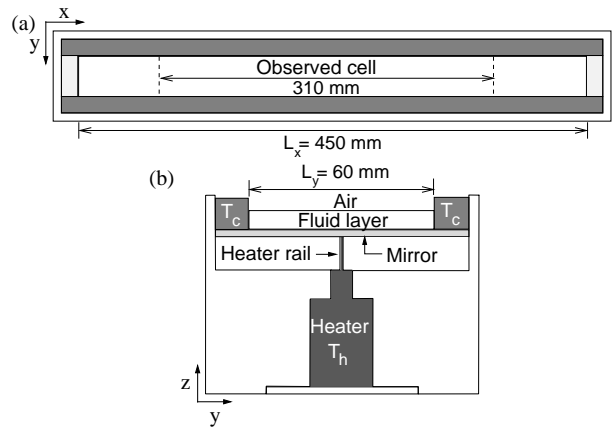}
\caption{\label{fig:1} Top view (a) and cross section (b) of the experimental cell. Both sketches do not keep the same scaling.}
\end{figure}

A fluid layer of depth $d$ is placed in a narrow rectangular cell $L_x \times L_y$ ($L_x = $ 470 mm, $L_y = $ 60 mm) [Fig.~\ref{fig:1}(a)]. This layer lies over a mirror [Fig.~\ref{fig:1}(b)] in order to implement the shadowgraphy technique. At the center and along a line in the $\hat x$ direction, the layer is heated at $T_h$ by means of a thermoregulated rail of 1 mm (thickness). This rail is placed beneath the mirror and it is laterally isolated providing a smoothed gaussian temperature profile in the $\hat y$ direction . The lateral walls are two aluminum blocks (coolers) whose temperatures are kept constant at $T_c = $ 20.0 $\pm$ 0.1 $^{\circ}$C by means of a secondary water circulation. A more detailed description of the experimental setup is found in~\cite{Miranda08}. \par

\begin{figure}
\includegraphics[width=8cm]{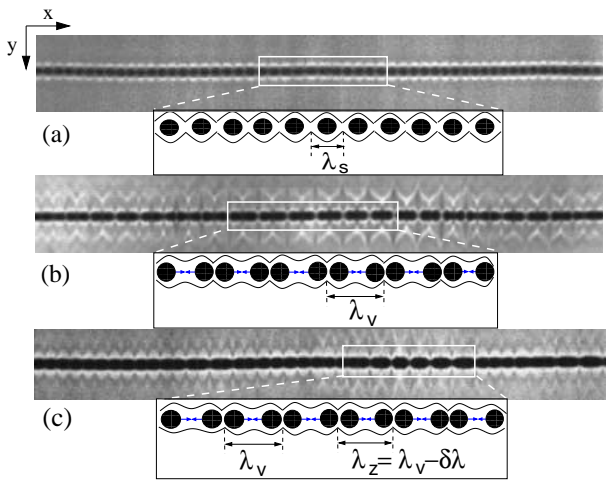}
\caption{\label{fig:2} Instantaneous shadowgraphies of the patterns: (a) ST for the stationary mode $M_s(k_s,0)$; (b) ST/ALT for the oscillatory domains with counter propagating modes $M_{v\pm}(k_{v},\pm\omega_v)$ which are resonant with $M_s$; (c) ST/ZZ for the spatiotemporal beating regime with new modes $M_{z\pm}(k_z,\omega_z)$ very close to the previous $M_{v\pm}$. Below we show a zoomed sketch of the oscillators involved and their characteristic wavelengths ($\lambda_s$, $\lambda_v$, $\lambda_z=\lambda_v-\delta\lambda)$.}
\end{figure}

The fluid used is silicone oil with a viscosity of 5 cSt. This fluid is transparent to visible light with a slow evaporation rate, therefore in the range of temperatures explored we can work with the free surface of the layer opened to the atmosphere whose temperature is controlled at $T_a = $ 22.0$\pm 0.1^{\circ}$C. The control parameter for a fixed depth is the vertical temperature difference, $\Delta T_v = T_h - T_a$. The relevant physical properties of the 5 cSt silicone oil are: the thermal conductivity $\lambda =$ 0.117 Wm$^{-1}$K$^{-1}$, the thermal diffusivity  $\kappa =$6.68$\: 10^{-8}$ m$^2$s$^{-1}$, and the surface tension $\sigma =$ 19.7 mN m$^{-1}$. \par

The results reported here are measured for a fixed depth $d =$ 3 mm. The most suitable adimensional number for the B\'enard-Marangoni convection mechanism is the dynamic Bond number $Bo_D = R/M $ ($\equiv${\sl Rayleigh number/ Marangoni number}), for which at $\Delta T_v =$ 30 K  we get $Bo_D \approx$ 1.04. In consequence, thermobuoyancy effects (R) are balanced by thermocapillary effects (M). There are two basic characteristic time scales: the viscosity time scale $\tau_\nu \approx$ 2 s and the thermal diffusivity time scale $\tau_\kappa \approx$ 135 s, hence the convective time scales are supposed to be subordinated to the diffusive temperature scales.\par

The shadowgraphic flow-visualization technique allows us to observe thermal gradients in the bulk of the fluid layer when an incoming parallel light is sent through the convection pattern. Once it is reflected back at the mirror surface, the output beam is projected into a screen. The space-time varying modulation of the light beam on the screen reveals the characteristic spatiotemporal periodicity of each pattern, in Fig.~\ref{fig:2} we show some images. The darker aligned spots in the center correspond to the uprising thermal plumes (or hotspots) at the heating line. The screen image is recorded with a CCD video camera with a resolution of 570 $\times$ 485 pixels and constant focal length, aperture and gain. The optical devices allow us to visualize on the screen a centered area of 310 mm $\times$ 60 mm from the total surface area (450 mm $\times$ 60 mm) [Fig.~\ref{fig:1}(a)].\par

From the image on the screen (Fig.~\ref{fig:2}) the acquisition image system records along a line two kinds of spatiotemporal diagrams: one over the bright modulation parallel to the aligned hotspots at a fixed contrast position $y_o$, $S_{y_o}(x,t)$ [Fig.~\ref{fig:3}(a,d)], and another one perpendicular to it at a given position $x_o$, $S_{x_o}(y,t)$ [Fig.~\ref{fig:4}(a)]. \par

The following results and theoretical discussions concern the study of a 1D-convective system whose geometric aspect ratios, at $d =$ 3 mm, are $\Gamma_x = L_x/d =150$ and $\Gamma_y = L_y/d = 20$, hence the ratio $\frac{\Gamma_x}{\Gamma_y}=$ 7.5 allows to classify the system as weakly confined in {\it x}.

\subsection*{Measurement process}

\begin{figure*}
\includegraphics[width=16cm]{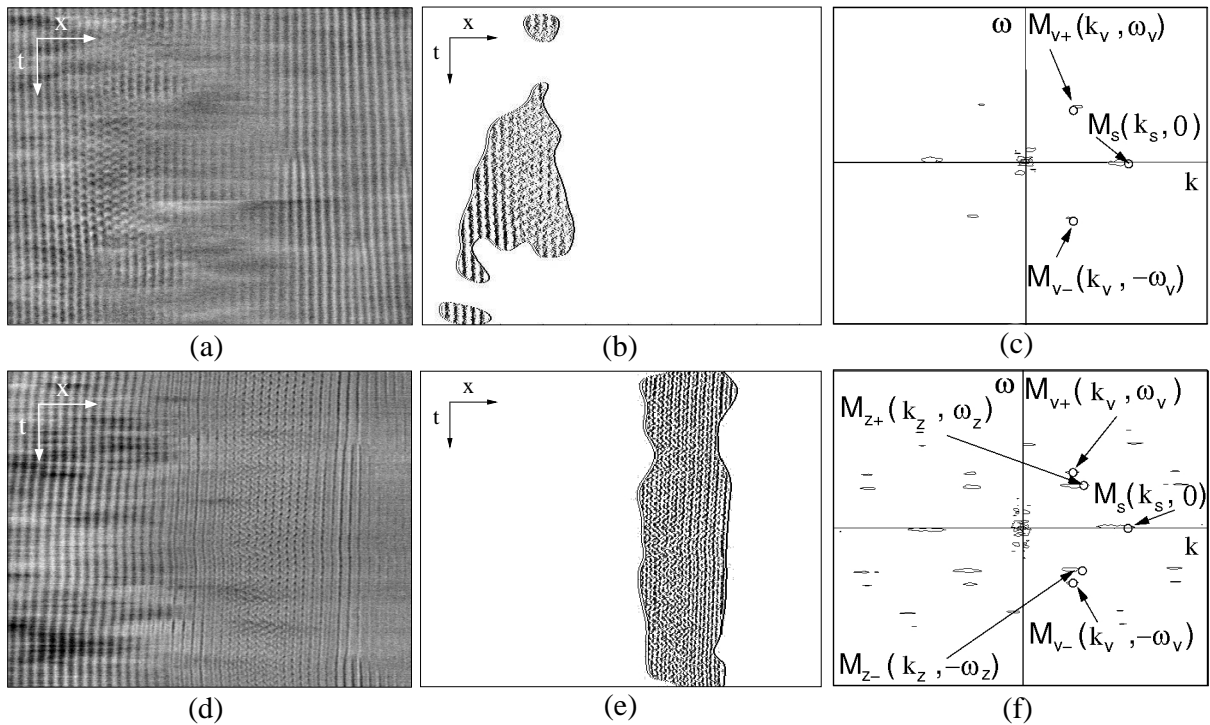}
\caption{\label{fig:3} Spatiotemporal diagrams $S_{y_o}(x,t)$ in the ST/ALT (a) and the ST/ZZ (d) regimes; (b) and (e) are the corresponding filtered signals by complex demodulation techniques with $\mu =1/e$; and (c) and (f) are the corresponding Fourier spectra where the fundamental modes have been defined.}
\end{figure*}

Spatiotemporal diagrams $S_{y_o}(x,t)$ are processed via the bidimensional Fourier transform and complex demodulation techniques in order to determine the modulus of the amplitudes $|A_{j}|$, the wave numbers $k_j$ and the frequencies $\omega_j$ of each fundamental Fourier j-mode $M(k_j,\omega_j)$ [see Fig.~\ref{fig:3}(c,f)]. So, each oscillating pattern will be defined by the following field: $\Psi (x,t)=\sum_j A_j \exp{\left[i\left(k_{j}x+ \omega_j t\right)\right]}+ c.c$. 
Also, demodulation techniques allow us to characterize the dynamics of localized patterns and fronts by defining the following parameters on $S_{y_o}(x,t)$:
\begin{itemize}
\item[(a)] The signal $S_{y_o}(x,t)$ is processed by dismissing the normalized amplitude above a critical value $\mu=1/e$ [see Fig.~\ref{fig:3}(b,e)]. This method is used to determine the average size of localized patterns, $L_c$, and the nature of their boundaries: stationary and fluctuating fronts. Thus, the parameter $L_c$ is a coherence length. A similar parameter is defined in other systems like in the K\"{u}ppers-Lorz instability~\cite{Tu92}, the plane Couette flow~\cite{ThBottin} and the Belousov-Zhabotinsky reaction~\cite{Marts04}.\par 
\item[(b)] The velocity of propagation of boundaries, $v_p$, is determined by measuring the velocity of the front defined by the new traveling mode with $\mu =1/e$. This method is not suitable for fluctuating boundaries.
\item[(c)] The invasion rate, $\sigma_{inv}$, is the ratio between the total surface occupied by the domains belonging to a certain mode with respect to the total extension of $S_{y_o}(x,t)$. This parameter is measured for the new unstable mode or modes invading the whole system.
\item[(d)] The attenuation length, $\xi$, is measured for coexisting patterns, and it is the length for which the new unstable pattern penetrates further beyond the coexisting domain~\cite{Ribotta79,Miranda08}. We choose the unstable mode which represents the new pattern, and we measure the length for which its amplitude decays according to $A \sim e^{-x/\xi}$ (from critical phenomena~\cite{Huang87}), where $\xi$ is a correlation length of the system. This parameter might be understood as a ``spatial memory'' inside the neighboring domain. This measurement is not suitable for fluctuating boundaries.

\end{itemize}

From the unidimensional Fourier transform of the signal $S_{x_o}(y,t)$, qualitative and quantitative information can be extracted about the dynamics of a particular hotspot. We have checked that outside the acquisition position selected to record $S_{y_o}(x,t)$, the temporal frequencies of a given oscillator are the same than the ones provided by $S_{y_o}(x,t)$. The sequences of measurements are taken for ascending steps of 1 and 0.5 K, and correspond to the asymptotic states.\par

\section{RESULTS AND DISCUSSION}

\subsection{A general approach to the experimental results}

\begin{figure}
\includegraphics[width=8cm]{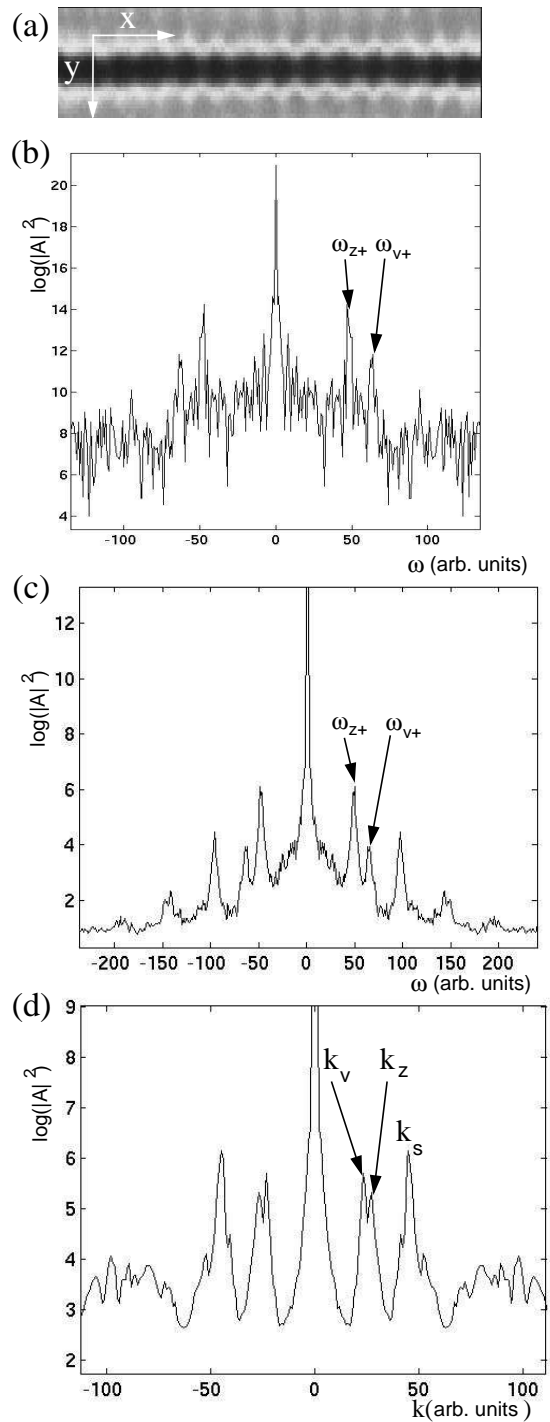}
\caption{\label{fig:4} In the ST/ZZ regime, at $\Delta T_v=$ 33.5 K : (a) Spatiotemporal diagram $S_{x_o}(y,t)$ for an oscillator belonging to the ZZ domain; (b) Fourier spectrum in $\omega$ from the signal $S_{x_o}(y,t)$ [in (a)] at the acquisition position to record $S_{y_o}(x,t)$. From the signal $S_{y_o}(x,t)$: (c) The Fourier spectrum in $\omega$ [the scaling is different from (b)]; (d) The Fourier spectrum in $k$. The components of the critical mode $M_{z+}(k_z,\omega_z)$ and the previous existing one $M_{v+}(k_{v},\omega_v)$ have been indicated in their respective graphics.}
\end{figure}

\begin{figure}
\includegraphics[width=8cm]{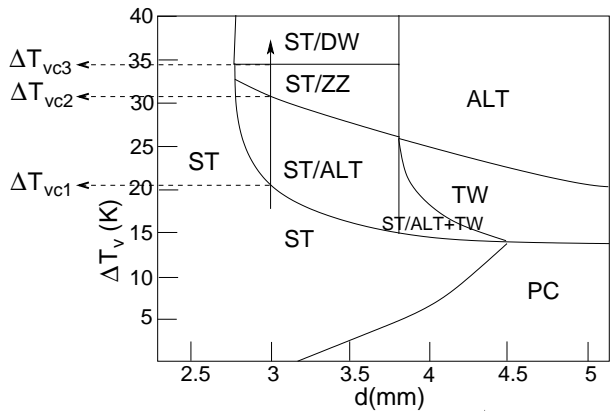}
\caption{\label{fig:5} Stability diagram. Continuous lines bound regions with the same asymptotic dynamics. Discontinuous arrows correspond to the critical points, the continuous upward arrow corresponds to the ascending sequences of measurements. Stationary patterns are: PC (primary convection) and ST (cellular pattern). Oscillatory patterns are: ST/ALT (mixed pattern of irregular clusters), ST/ZZ (spatiotemporal beating regime of stationary clusters), ST/DW (temporal beating regime) and TW (traveling waves). ST/ALT+TW corresponds to the coexisting ST/ALT and TW patterns}. 
\end{figure}

\begin{figure*}
\includegraphics[width=16cm]{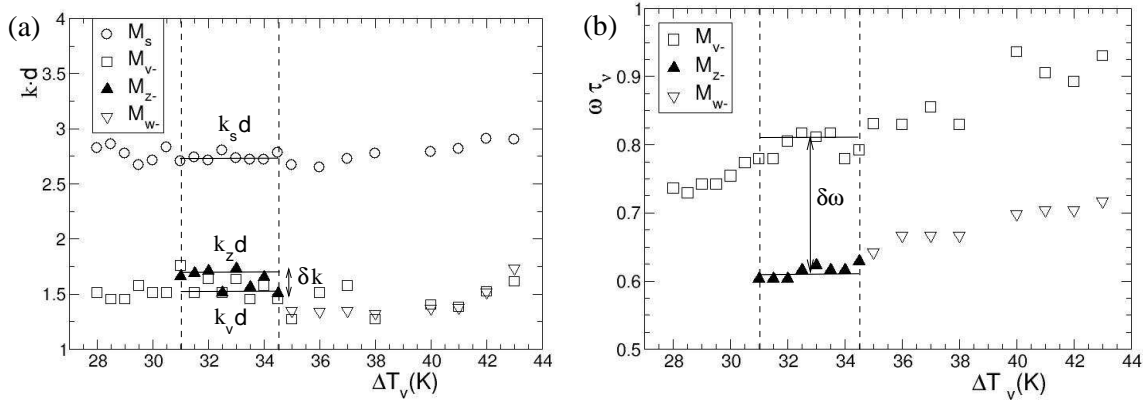}
\caption{\label{fig:6} Evolution of the wavenumbers (a) and the frequencies (b), respectively scaled with $d=$ 3 mm and $\tau_{\nu}=d^2/\nu=$ 1.8 s. Vertical dashed lines correspond to threshold values ($\Delta T_{vc2}=$ 31 K, $\Delta T_{vc3}=$ 34.5 K) which bound the ST/ZZ regime and continuous lines in the ST/ZZ regime are a guide to the eye.}
\end{figure*}

\begin{figure*}
\includegraphics[width=16cm]{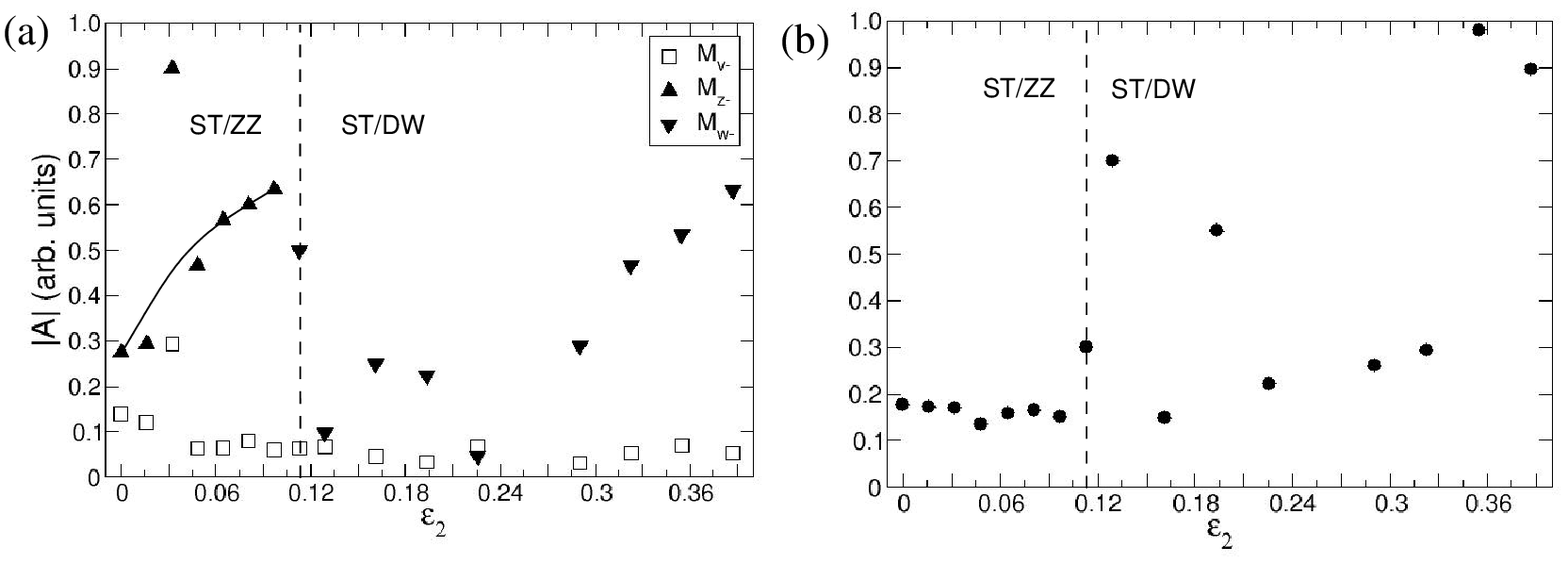}
\caption{\label{fig:7} (a) Evolution of the amplitudes of the fundamental modes $M_{z-}$, $M_{v-}$ inside the domains defined by ZZ and DW; (b) Evolution of the amplitude of the stationary mode $M_s$ inside the same domains in order to filter the resonant contribution of the mode $M_s$ at the boundaries [the scaling in amplitudes is different from (a)]. Dashed lines correspond to threshold values and continuous lines are a guide to the eye. In (a) and (b) $\varepsilon_2=\Delta T_v/\Delta T_{vc2}-1$.}
\end{figure*}

In the bulk of the fluid layer, for every control parameter, a primary convection (PC) appears. It consists on two counterpropagative rolls which ascend at the heating line and descend at the lateral cooled walls. As soon as $\Delta T_v\not= 0$, a primary bifurcation to the stationary ST pattern takes place (see the stability diagram in Fig.~\ref{fig:5}). In the shadowgraphy image of ST [Fig.~\ref{fig:2}(a)] we observe this multicellular pattern, with wavelength $\lambda_{s}\approx$ 6 mm [$\lambda_{s} \approx 2d$, see Fig.~\ref{fig:6}(a)]. The physical aspect ratio defined as $\Gamma_{\lambda_{s}}=L_x/\lambda_{s}\approx 80$ is large enough and thus, under this condition the system is extended. The ST pattern is represented by the stationary mode $M_s(k_s,0)$. At $d=$ 3 mm this pattern never looses stability against the emerging unstable traveling modes. This is the cause for which we always find localized domains of the new pattern coexisting with the ST pattern. As we increase the control parameter $\Delta T_v$, the transition to weak turbulence starts. For the different regimes, results on $k$ and $\omega$ (in Fig.~\ref{fig:6}) and on modules $|A|$ (in Fig.~\ref{fig:7}) are obtained from the demodulation analysis of the spatiotemporal diagrams $S_{y_o}(x,t)$. The subsequent secondary instabilities give rise to the following asymptotic patterns summarized in table~\ref{tab:1} (see also the stability diagram in Fig.~\ref{fig:5}): 

\begin{itemize}
\item[(i)] A mixed ST/ALT pattern. At $\Delta T_{vc1}=$ 21 K, the system bifurcates towards a regime of fluctuating domains, these are irregular clusters of variable size in the ALT pattern that usually collapse [Fig.~\ref{fig:3}(a,b)]. This ALT domains show a period doubling of the basic pattern ($k_{v}=k_s/2$), with two counterpropagative modes $M_{v\pm}(k_{v},\pm \omega_v)$ superimposed to the stationary mode $M_s(k_s,0)$. According to the weakly nonlinear coupling revealed from the Fourier spectra [i.e. Fig.~\ref{fig:3}(c)], inside the fluctuating fronts the two traveling modes locally destabilize the basic ST pattern, so the resonant triad [$M_s(k_s,0)$, $M_{v\pm}(k_{v},\pm \omega_v)$] is triggered. 
According to results in Fig.~\ref{fig:6}(b) traveling modes have an average period of the order of 15 s. 

\item[(ii)] A spatiotemporal beating regime or ST/ZZ pattern. At $\Delta T_{vc2}=$ 31 K, and for a range of 5 K, one or two large domains in ZZ, with an average width of $L_c\approx$ 80 mm coexist with the ST pattern [see the coherent ZZ domain in Fig.~\ref{fig:3}(d,e)]. The ZZ domains are stationary clusters resulting from the splitting of the spatiotemporal frequencies of the fundamental modes. The spatial beating phenomenon comes from the existence of two close wavenumbers ($k_z$, $k_{v}$) of the traveling modes [Fig.~\ref{fig:4}(d)], meanwhile their characteristic zig-zag geometry is due to the temporal beats [two close temporal frequencies ($\omega_v$, $\omega_z$) extracted identically from diagrams $S_{x_o}(y,t)$ in Fig.~\ref{fig:4}(b) and diagrams $S_{y_o}(x,t)$ in Fig.~\ref{fig:4}(c)]. 
From the bidimensional Fourier spectrum [Fig.~\ref{fig:3}(f)] we confirm the increasing number of harmonic modes taking part in a stronger nonlinear dynamics. So, this beating phenomenon corresponds to the following fundamental modes: the stationary mode $M_s(k_s,0)$, the new emerging traveling modes $M_{z\pm}(k_z,\pm \omega_z)$, and the previous existing ones $M_{v\pm}(k_{v},\pm \omega_v)$. From the results on Fig.~\ref{fig:6}(a,b), we obtain a narrow spatiotemporal frequency band $(\delta k,\delta \omega)$, where $\delta k = |k_z-k_{v}|$ and  $\delta \omega = |\omega_v-\omega_z|$ are the envelope periodicities of the spatial and temporal beats, respectively. 

\item[(iii)] A temporal beating regime or ST/DW pattern. From $\Delta T_{vc3}=$ 34.5 K [Fig.~\ref{fig:6}(a,b)] onward, the spatial frequency splitting disappears. So, in this regime the competition between the split traveling modes is not able to sustain spatial beats, only temporal beats remain by means of a new mode which becomes unstable in a unique and larger localized domain. This domain is composed of the following fundamental modes: $M_s(k_s,0)$, $M_{v\pm}(k_{v},\pm \omega_v)$, $M_{w\pm}(k_{v},\pm \omega_z)$.
\end{itemize}

In Fig.~\ref{fig:2}, we show the characteristic distribution of hotspots for these regimes: ST, ST/ALT and ST/ZZ. Discontinuities in the dynamics of hotspots denote the presence of 1D-fronts. We may think of hotspots as a sort of ``coarse-grained units'' linked to the hydrodynamical variables which are able to oscillate like limit-cycle oscillators. Each characteristic pattern depends on the couplings between oscillators, future work is devoted to determine different types of couplings, this topic is discussed in the last section. A review of patterns for thicker layers is found in~\cite{Miranda08}.\par

We have found no hysteresis for neither of these instabilities, i.e. there is no different behavior in the descending sequence compared with the ascending one presented here. For higher depths, classical subcritical bifurcations have been found~\cite{Miranda08}. From the diagrams $S_{y_o}(x,t)$, in Fig.~\ref{fig:7}(a,b) we show the evolution of the modulus of the amplitude of each fundamental mode belonging to the same quadrant of the symmetric Fourier spectrum: $M_s$, $M_{v-}$, $M_{z-}$. In order to avoid large values of $|A_s|$ due to the resonance of the stationary mode at the boundaries~\cite{Miranda08}, data in Fig.~\ref{fig:7}(b) is determined inside the boundaries confining ZZ and DW domains.\par

When we disturb the surface locally and close to the threshold of the bifurcations to ST/ALT (at $\Delta T_{vc1}$) and ST/ZZ (at $\Delta T_{vc2}$), the corresponding oscillatory domains become globally unstable in transient regimes. Besides, fronts show no transversal drifting, so the group velocity is null and both instabilities have an absolute character. \par

\begin{table*}
\caption{\label{tab:1}Regimes in the route to spatiotemporal chaos for increasing $\Delta T_v$}
\begin{ruledtabular}
\begin{tabular}{lllllll}
\noalign{\smallskip}
Stationary pattern ({\bf ST}) & extended pattern & 1 mode: $M_s(k_s,0)$\\
Mixed pattern ({\bf ST/ALT}) & localized domains & 3 modes: $M_s(k_s,0)$, $M_{v\pm}(k_{v},\pm \omega_v)$\\
Spatiotemporal beats ({\bf ST/ZZ}) & localized domains & 5 modes: $M_s(k_s,0)$, $M_{v\pm}(k_v,\pm\omega_v)$, $M_{z\pm}(k_z,\pm\omega_z)$\\
Temporal beats ({\bf ST/DW}) & localized domains & 5 modes: $M_s(k_s,0)$, $M_{v\pm}(k_v,\pm\omega_v)$, $M_{w\pm}(k_v,\pm\omega_z)$\\
\noalign{\smallskip}
\end{tabular}
\end{ruledtabular}
\end{table*}

For consistency with the following sections, we define two reduced control parameters $\varepsilon_1=\Delta T_v/\Delta T_{vc1}-1$ and $\varepsilon_2=\Delta T_v/\Delta T_{vc2}-1$ for the two subsequent instabilities towards the mixed ST/ALT pattern and towards the spatiotemporal beating regime ST/ZZ, respectively.\par

\subsection{Irregular clustering dynamics in the mixed pattern}

\begin{figure*}
\includegraphics[width=16cm]{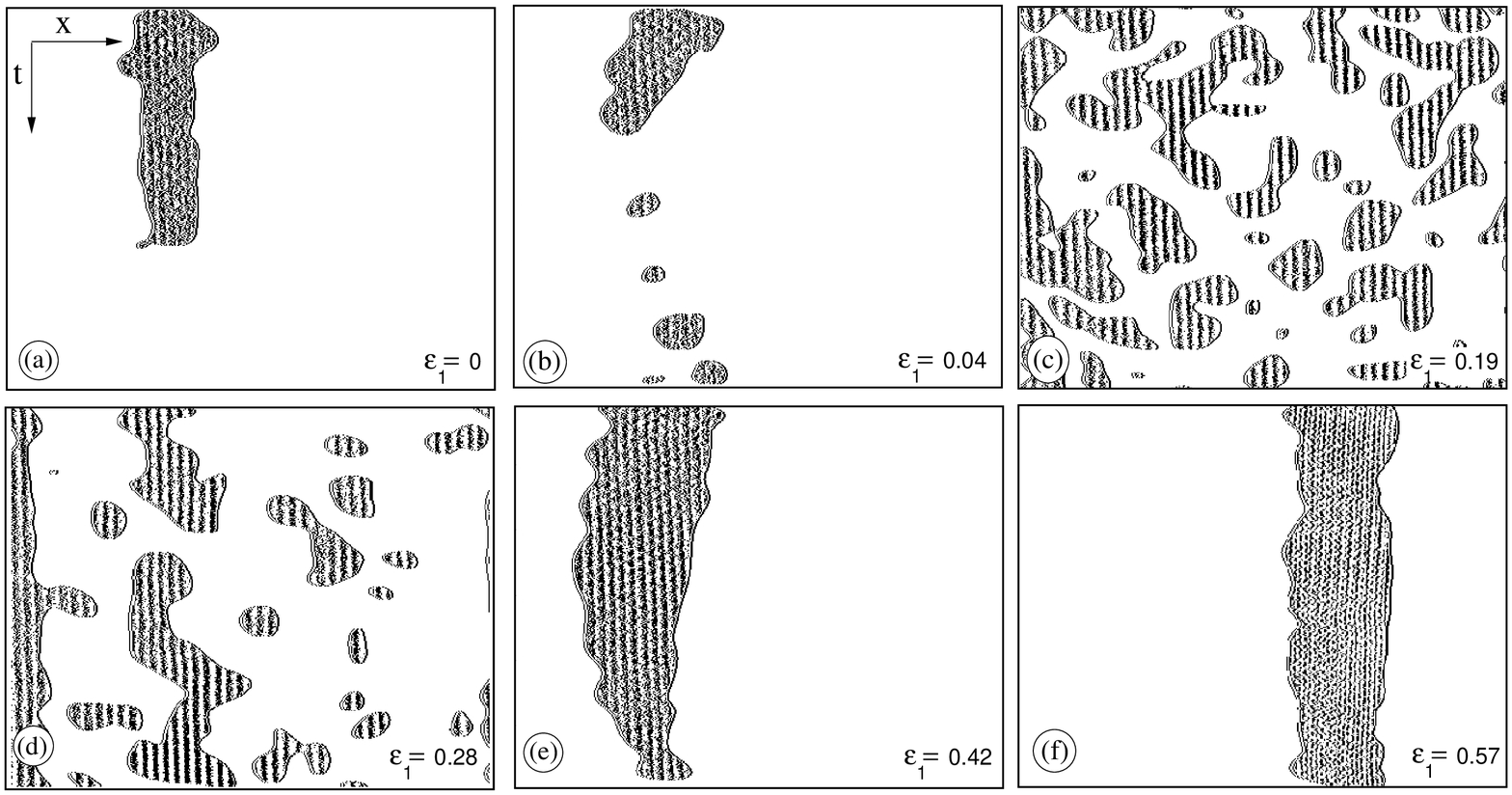}
\caption{\label{fig:8} Sequence of filtered diagrams $S_{y_o}(x,t)$ with $\mu=1/e$, the reduced control parameter is defined as $\varepsilon_1=\Delta T_v/\Delta T_{vc1} - 1$ with $\Delta T_{vc1}=$ 21 K. Diagrams (a)-(e) correspond to the mixed ST/ALT pattern, diagram (f) corresponds to the beating regime ST/ZZ at the threshold ($\varepsilon_2=0$).}
\end{figure*}

In Fig.~\ref{fig:8} we show a sequence of filtered spatiotemporal diagrams in the mixed ST/ALT regime where the surrounding ST pattern has been filtered. The system is bistable in this regime because localized domains in ALT, with well defined and fluctuating boundaries, coexist with the basic cellular pattern ST. Fluctuating boundaries are irregular fronts which bound these localized domains, hotspots belonging to these domains are synchronized in $\omega_v$ and therefore the images in Fig.~\ref{fig:8}(a-d) remind us the nature of irregular clusters (eventually they collapse) like the ones appearing in the Belousov-Zhabotinsky reaction-diffusion systems~\cite{Vanag00}. 
Above the threshold ($\varepsilon_1=0$) clusters spread over the whole diagram $S_{y_o}(x,t)$ [i.e. Fig.~\ref{fig:8}(c)] until they stop collapsing at $\Delta T_v \approx$ 26 K ($\varepsilon_1=0.24$), and saturate into a unique domain of great spatiotemporal coherence [i.e. Fig.~\ref{fig:8}(e)]. This domain will remain in the following instability to spatiotemporal beats [i.e. Fig.~\ref{fig:8}(f)] where two new unstable modes will emerge. \par

\begin{figure}
\includegraphics[width=8cm]{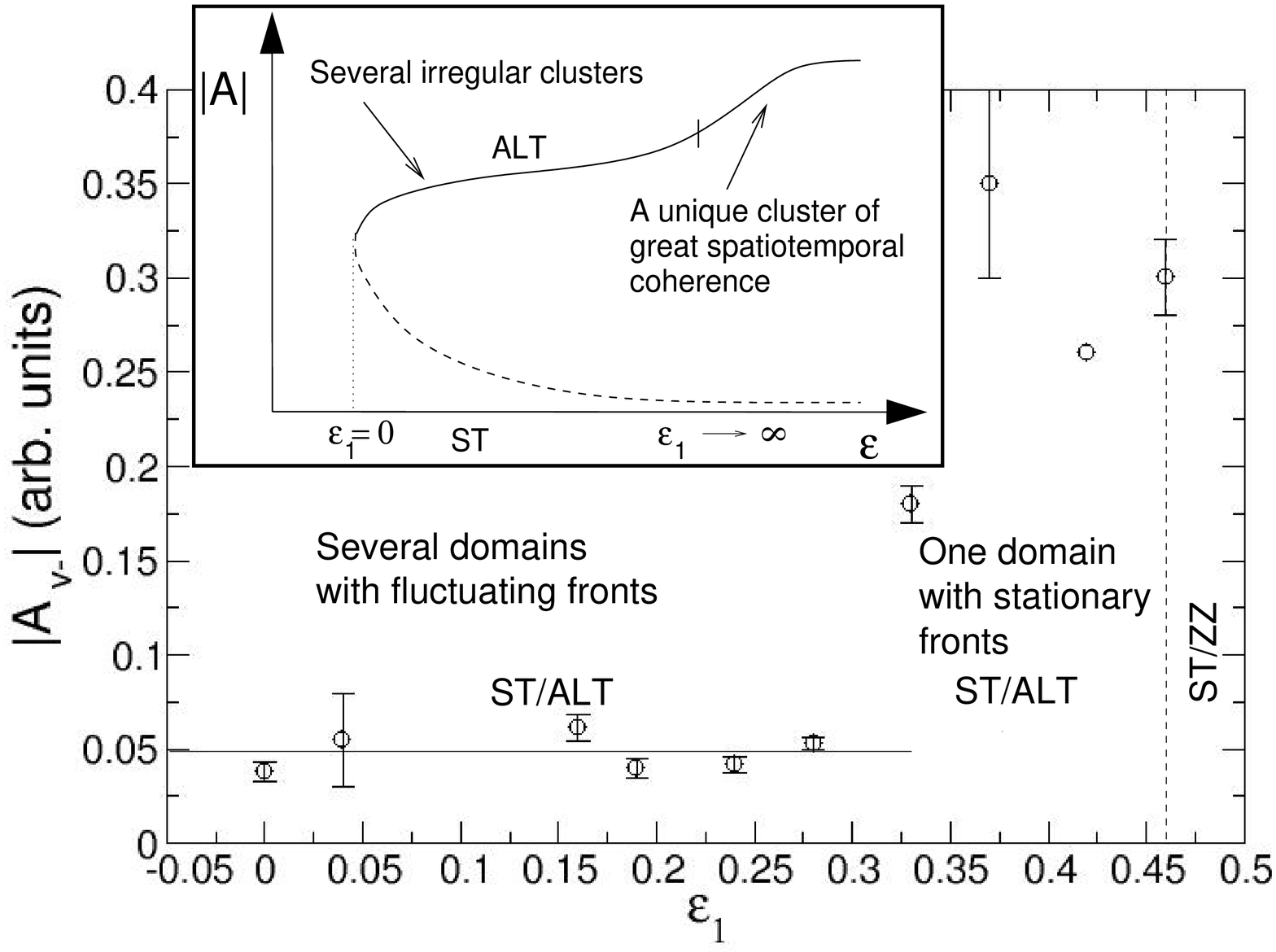}
\caption{\label{fig:9} Evolution of the amplitude of the fundamental traveling mode $M_{v-}$ along an ascending sequence of $\varepsilon_1=\Delta T_v/\Delta T_{vc1}-1$ inside the bifurcated domains. Vertical dashed lines separate different regimes and the continuous line corresponds to the averaged amplitudes. The inserted sketch corresponds to the diagram of bifurcation to the ST/ALT regime from the stationary ST regime.}
\end{figure}

The basic cellular pattern ST does not ever become unstable in the wide range of the explored control parameters ($\Delta T_v = [21, 43]$ K). According to previous experimental work (see references in ~\cite{Saarloos03}), the bifurcation diagram would correspond to that of a global subcritical bifurcation where the threshold branch is unstable for a diverging control parameter $\varepsilon_1$ (see sketch in Fig.~\ref{fig:9}). If we take into account thermocapillary effects competing against thermogravitatory effects, from the dynamic Bond number $Bo_D\propto d^2$, we infer that by decreasing $d$, thermocapillary effects become stronger in the mixed ST/ALT regime than for higher values of $d$, where this dynamical regime coexists with a traveling wave pattern~\cite{Miranda08} (see the stability diagram Fig.~\ref{fig:5}). Consequently, we may consider that at $d=$ 3 mm subcriticality is sent to infinity, and bistability is a natural consequence in the following instabilities where critical modes become unstable inside these confined domains. However, the subsequent instability might not necessarily be subcritical as it will be shown in the next section.\par

\begin{figure*}
\includegraphics[width=16cm]{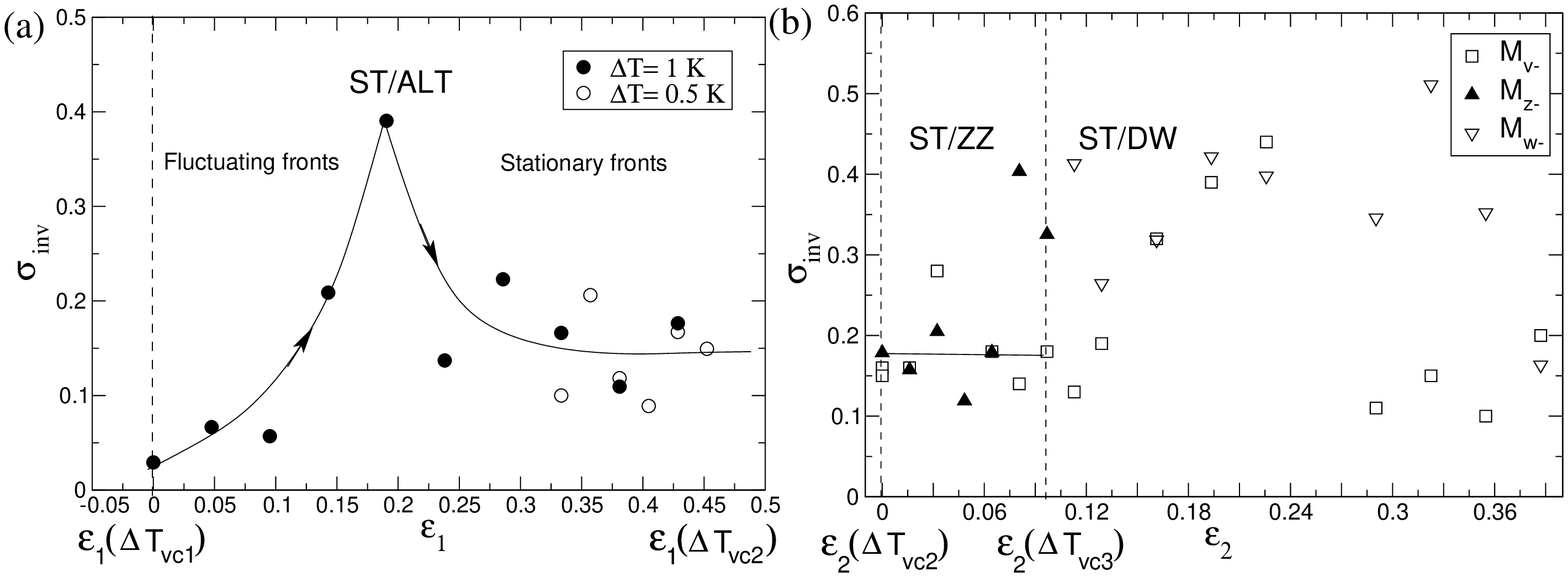}
\caption{\label{fig:10} (a) Invasion rate of the traveling mode $M_{v-}$ close and above the bifurcation to the mixed ST/ALT pattern for two different sequences by steps $\Delta T=1$ and 0.5 K. The reduced control parameter is $\varepsilon_{1}= \Delta T_v/\Delta T_{vc1} -1$. Continuous lines are a guide to the eye; (b) Invasion rate determined by the modes $M_{z-}$ and $M_{v-}$ representing the beating regime with $\mu=1/e$ and along an ascending sequence. The reduced control parameter is  $\varepsilon_{2}= \Delta T_v/\Delta T_{vc2} -1$. Vertical dashed lines separate different regimes at threshold values: $\Delta T_{vc1}=$ 21 K, $\Delta T_{vc2}=$ 31 K, $\Delta T_{vc3}=$ 34.5 K.}
\end{figure*}

In this transition, the amplitudes of the traveling modes are constant (see Fig.~\ref{fig:9} where $|A_{v+}| \approx |A_{v-}|$) in the range $0<\varepsilon_1<0.35$ ($\varepsilon_1=\Delta T_v/\Delta T_{vc1}-1$). The frequency $\omega_{v}$ increases monotonically with $\Delta T_v$ and has an average value of 0.42 s$^{-1}$ ($<\omega\tau_{\nu}> =$ 0.75) [Fig.~\ref{fig:6}(b)].
In Fig.~\ref{fig:10}(a) we show the invasion rate of clusters $\sigma_{inv}$ $vs.$ $\varepsilon_1$. Following the sequence of pictures in Fig.~\ref{fig:8} we infer that, from an initial domain in ALT, as we increase the control parameter ($\varepsilon_1$), $\sigma_{inv}$ increases until it spreads inhomogeneously ($\sigma_{inv}\approx 0.4$)  all over the spatiotemporal diagram $S_{y_o}(x,t)$. This process corresponds to the left branch in Fig.~\ref{fig:10}(a) with fluctuating fronts. Similarly, in other transitions to turbulence in open flows we may say that, at this stage, the system is being ``contaminated'' (i.e. plane Couette flow \cite{Bottin98}). Once irregular clusters are equally distributed over the spatiotemporal diagram, the average size of these irregular clusters is $L_c \approx 30$ mm.  The spatial and temporal correlation lengths of the stationary mode decay in this regime. This is the cause for considering ST/ALT as a spatiotemporal chaos regime.\par

From this point onward, $\sigma_{inv}$ decreases until it achieves a practically constant value of approximately 0.14. This is due to the fact that we are settling the system close and below the following instability with the formation of a unique ALT domain of great spatiotemporal coherence and with stationary fronts ($v_p\approx$ 0.01 mm/s). This behavior corresponds to the right branch in Fig.~\ref{fig:10}(a). We have checked the same behavior below the following bifurcation with a smaller step. Regarding these results there is a growth of the ``phase coherence'' which is supposed to fit a similar role as the invasion rate when it increases in the left branch. In this left branch, as we increase the control parameter more oscillators are synchronized nearby $\omega_v$ according to an irregular clustering dynamics.\par

The theoretical approach to this clustering process is close to the diffusion-induced inhomogeneity proposed by H. Daido and K. Nakanishi~\cite{Daido07} except for the global coupling, if we suppose that hotspots are being locally coupled by diffusion and convection. The growth of phase coherence is supposed to trigger the clustering process until the ``phase synchronization'' spreads over the whole array of oscillators, when there are two stationary clusters (ST/ZZ regime).\par 
 
\subsection{Supercritical bifurcation to the stationary clusters regime}

\begin{figure*}
\includegraphics[width=10cm]{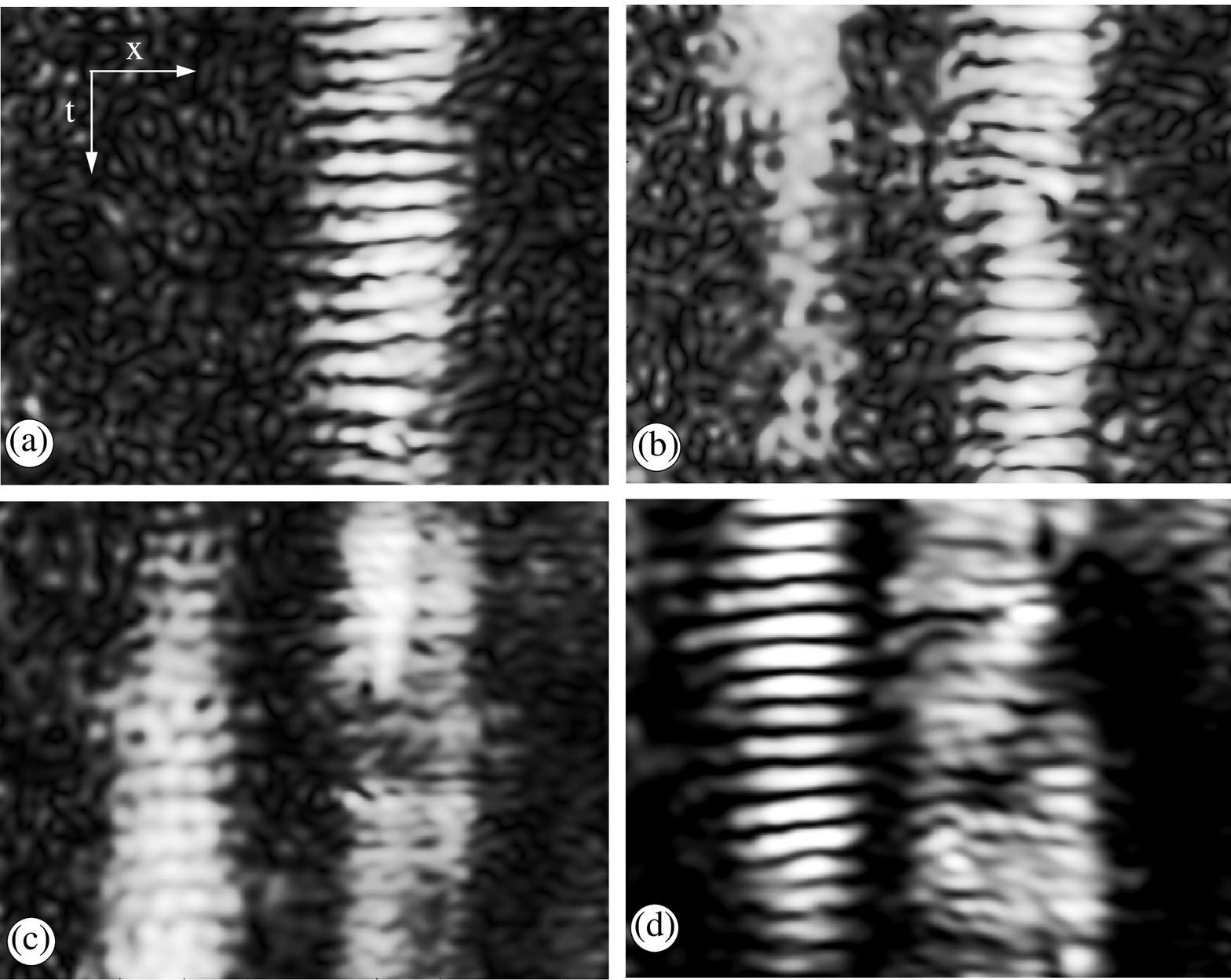}
\caption{\label{fig:11} A sequence of demodulated diagrams $S_{y_o}(x,t)$ selecting the pair $M_{v-}$, $M_{z-}$ for increasing $\Delta T_v$ in the ST/ZZ regime: (a) 31 K ($\varepsilon_2=$ 0), (b) 31.5 K ($\varepsilon_2=$ 0.02), (c) 32 K ($\varepsilon_2=$ 0.03), (d) 33.5 K ($\varepsilon_2=$ 0.08). Dark and bright regions correspond to the minimum and maximum values of the amplitude.}
\end{figure*}

For even higher temperatures the system becomes unstable towards the beating regime. In this ST/ZZ pattern, studied along two ascending sequences by steps of 0.5 and 1 K, two domains saturate to a fixed width value of $L_c\approx 80$ mm. The ALT pattern inside has become unstable because of the emerging unstable modes. These are two new traveling modes, $M_{z\pm}$, which are very close to the previous existing ones, $M_{v\pm}$. From Fig.~\ref{fig:6}(a,b) we extract, in the ST/ZZ regime, the envelope frequencies: $\delta k \approx$ 0.04 mm$^{-1}$ and $\delta \omega \approx$ 0.10 s$^{-1}$. These results are in agreement with the spatiotemporal periodicities of the demodulated signals shown in Fig.~\ref{fig:11}(a-d), the spatial periodicity of the beats is of 157~mm ($\approx 2L_c$) and the time periodicity is of 124 s ($\approx$ 62 s of maximum amplitude, brightness periodicity of zig-zags). 
The beating dynamics of the ST/ZZ pattern depends on the competition between the unstable modes $M_{z\pm}$ and $M_{v\pm}$, their amplitudes and the spatiotemporal frequency band $(\delta k, \delta \omega)$. The invasion rate is approximately constant $\sigma_{inv}=$ 0.18 [Fig.~\ref{fig:10}(b)].\par

\begin{figure*}
\includegraphics[width=12cm]{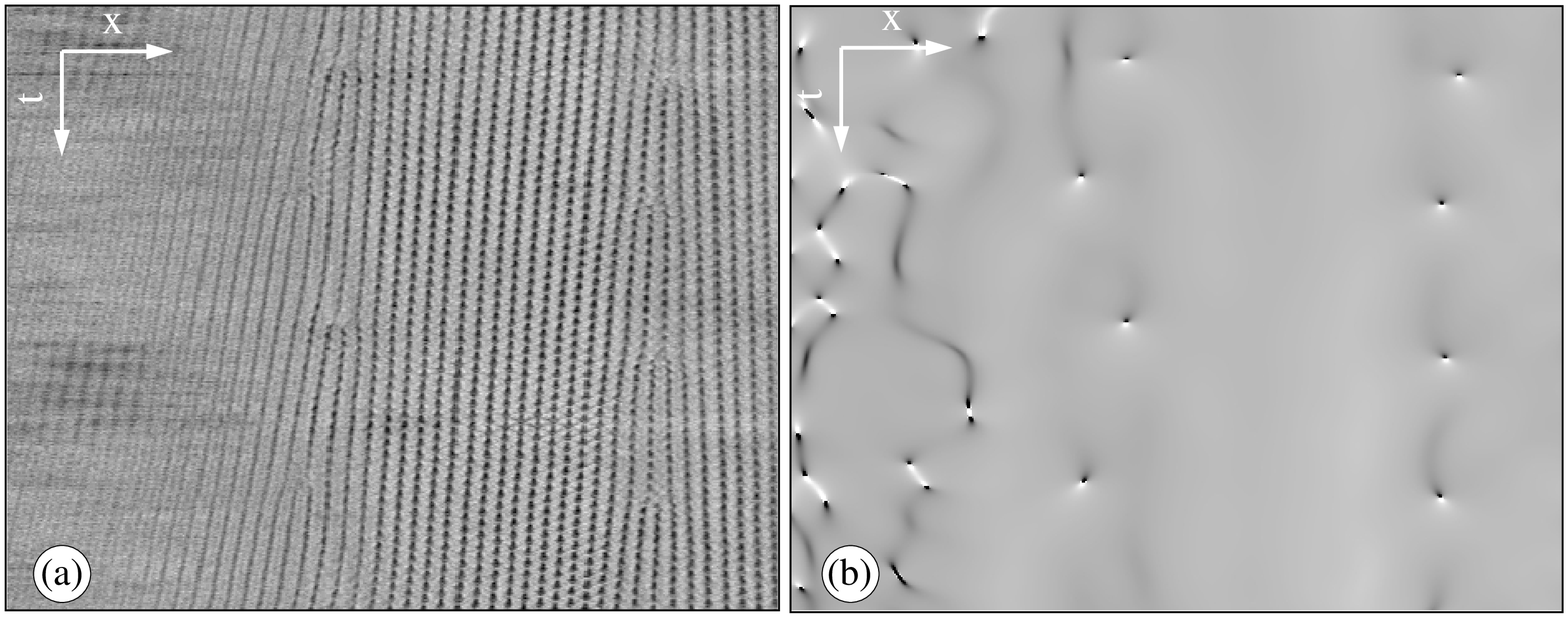}
\caption{\label{fig:12} (a) Spatiotemporal diagram $S_{y_o}(x,t)$ at $\Delta T_v=$ 42 K ($\varepsilon_2=$ 0.32); (b) the corresponding diagram for the phase gradient of the mode $M_s$, fronts of dislocations are observed.}
\end{figure*}

Above the threshold ($\varepsilon_2=$ 0) this regime is strongly nonlinear and beyond this instability, the pattern looses the spatial frequency splitting. Above $\Delta T_{vc3}\approx$ 34.5 K ($\varepsilon_2=0.11$), the temporal frequency splitting still remains in the new ST/DW pattern [Fig.~\ref{fig:7}(a,b)]. In this ST/DW pattern, a unique and wider domain in DW exists whose 1D-fronts, as we increase $\Delta T_v$, convert into a chain of dislocations [i.e. Fig.~\ref{fig:12}(a,b)]. Besides the stationary mode $M_s$ drifts as it has already been observed in other systems for high control parameter values~\cite{deBruyn96}. Inside this DW domain we observe an increasing number of temporal beats (their period diminishes) the further the system is from the threshold, in agreement with the results in Fig.~\ref{fig:6}(b) where the distance $\delta \omega$ becomes larger.
Further above $\varepsilon_2\ge$ 0.29 a different dynamics due to the increasing number of dislocations is responsible for a variable $\sigma_{inv}$ [Fig.~\ref{fig:10}(b)]. In Fig.~\ref{fig:7}(a) we notice that the evolution of the rate of the amplitudes $|A_{z-}|/|A_{v-}|$ increases continuously from 2 up to 7 until the necessary condition (in $k$) for the existence of spatial beats does not accomplish at the ST/DW regime.\par

\begin{figure*}
\includegraphics[width=16cm]{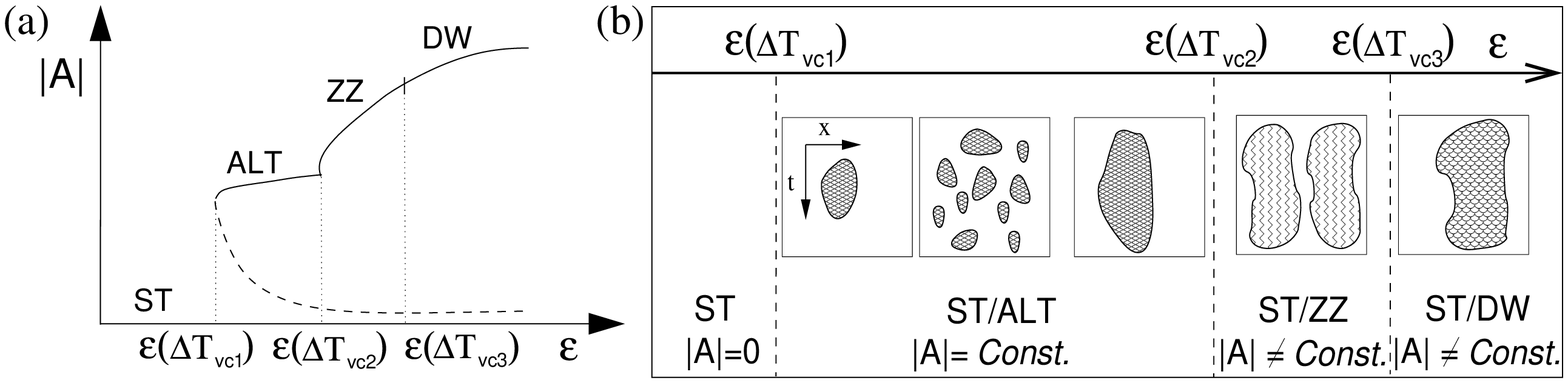}
\caption{\label{fig:13} (a) Bifurcation diagram $|A|$ $vs.$ $\varepsilon$, where  $\varepsilon=\Delta T_v/0.1$K $- 1$ is the reduced control parameter defined from the threshold of the instability to ST while $\varepsilon(\Delta T_{vc1})$, $\varepsilon(\Delta T_{vc2})$ and $\varepsilon(\Delta T_{vc3})$ are the reduced control parameter at the thresholds of the subsequent instabilities; (b) Sketch of the dynamics of the localized domains for the same increasing control parameter $\varepsilon$, the order parameter is $|A|$ for the corresponding critical modes. }
\end{figure*}

The continuous growth of the modulus of the amplitude of the critical mode $M_{z-}$ has been checked analyzing both the whole diagram $S_{y_o}(x,t)$ and the ZZ domain [Fig.~\ref{fig:7}(a)]. This continuous evolution of the modulus of the critical amplitudes $|A_{z\pm}|$ agrees with a supercritical bifurcation. On the other hand, from results in Fig.~\ref{fig:6} it follows that the phase velocity  $v_{\phi}= \omega/k$ along the whole cascade grows linearly with the vertical temperature gradient.\par

The supercritical transition to ST/ZZ is the beginning of a phase synchronization transition nearby the stationary cluster phase determined by the critical frequency $\omega_v$. This phase transition has to be completed far from the threshold $\varepsilon_2=0$~\cite{Miranda208}.
According to subcriticality (the subcritical ST branch is imposing bistability), from the analysis of the correlation length as an attenuation length of the new pattern (ST/ZZ) inside the original one (ST), we get: $<\xi>\approx 25$ mm. This subcritical length cannot be determined with fluctuating boundaries, nevertheless for the mixed ST/ALT pattern the role of $\sigma_{inv}$ has contributed to the interpretation of a sort of diverging correlation length that follows the inhomogeneous growth of the phase synchronization over the whole array of oscillators [Fig.~\ref{fig:10}(a)]. Regarding the supercriticallity of the critical modes in the ST/ZZ regime $M_{z\pm}(k_z,\pm \omega_z)$, the correlation length $\xi$ at the threshold is supposed to diverge, so its comprehension goes beyond the point of view of critical phenomena regarding the attenuation of critical amplitudes.\par

\section{FURTHER REMARKS AND CONCLUSIONS}

For a thermoconvective fluid layer, we have reported a route to spatiotemporal chaos for a system of high-dimensionality with N~$\approx$ 80 coupled oscillators. Synchronized interactions between hotspots define domains with the same oscillatory pattern and with specific 1D-fronts. These 1D-fronts can be of the fluctuating type, for irregular clusters (with weak coupling between oscillators), which take part in a process of inhomogeneous growth of the phase coherence. For higher control parameter values a new type of clusters emerge with stronger coupling between oscillators: 1D-fronts are stationary in this phase synchronization transition to ST/ZZ. Bifurcations to the mixed ST/ALT (irregular clusters) and ST/ZZ (stationary clusters) patterns are global instabilities where synchronized clusters have different coherent widths expressed in terms of $L_c$. We have shown that the size of the clusters at the beating regime is given by the wavelength of the spatial phase envelope and comes from a spatial frequency splitting. \par

In order to follow the whole sequence of bifurcations presented here, if the primary instability from PC towards the stationary ST pattern is produced at $\Delta T_v= 0.1$ K, we may define the reduced control parameter $\varepsilon=\Delta T_v/0.1$K $-1$.  The bifurcation diagram in Fig.~\ref{fig:13}(a) shows the cascade of bifurcations in the transition to spatiotemporal chaos for this particular extended system. Using the amplitude of the critical modes, as the most convenient order parameter, we show how the system undergoes a supercritical bifurcation to the beating regime ST/ZZ. There is a subcritical width in the ST/ZZ pattern determined by $L_c$ which is compatible with the existence of a supercritical bifurcation of the unstable traveling modes $M_{z\pm}$. This instability appears from a mixed ST/ALT pattern of localized clusters which are inhomogeneously spread. Oscillators belonging to this spatiotemporal chaos regime are synchronized in $\omega_v$, but often these irregular clusters collapse, similarly to an aging clustering dynamics for weak coupling. The growth of phase coherence accounts for the clustering process until the ensemble of N-coupled oscillators becomes phase synchronized in $\omega_z$ far from the threshold of this beating regime ST/ZZ ($\Delta T_{v2}$)~\cite{Miranda208}. Beyond this ST/ZZ pattern the system keeps only temporal beats, ST/DW. We have sketched the evolution of the clustering process in Fig.~\ref{fig:13}(b) where the order parameter is the module of the amplitude of the critical traveling modes, $|A|$. The dynamics in  the ST/ZZ regime depends on the competition between the split traveling modes.\par

The existence of localized domains in the beating regimes (ST/ZZ and ST/DW) approaches the statistical model for chaos of Hohenberg and Shraiman~\cite{Hohenberg89} in extended 1D-systems, when the scales of energy supply and dissipation are of the same order: the depth of the fluid layer $d$ (being $d\ll L_x$). Under these conditions, at $d =$ 3 mm our system exhibits the existence of two spatiotemporal coherent domains of width $L_c< L_x$. $L_c>\xi$, where $\xi$ has been defined as a correlation length of the system and measured as an attenuation length of the new ST/ZZ pattern inside the surrounding pattern ST ($<\xi> \approx$ 30 mm and $\xi \ll L_x$). In this context, the domains of width $L_c$ are correlated surfaces in space and time. Nevertheless, if we keep in mind the whole dynamics (from the mixed ST/ALT pattern to the beating regime) and look for a physical explanation to these kind of transition phenomena, we are pushed to understand synchronization transitions in a convective system.  For a given ensemble of N-limit cycle oscillators distributed on an 1D-array of extension $L_x$, depending on the type of diffusive coupling, the interaction range between oscillators, $\chi$, can be: (i) local (between neighboring oscillators, $\chi=L_x/(N -1)$) leading to spatiotemporal chaos; (ii) global (all-to-all, $\chi=L_x$) leading to collective synchronization phenomena; and (iii) nonlocal (further than the nearest neighbors, $L_x/(N-1) < \chi < L_x $) leading to the existence of coherent patterns. Cooperative phenomena are present in both secondary bifurcations, the interaction between hotspots corresponds to a diffusive coupling between nearest neighbors which, as far as we increase the control parameter, becomes stronger due to nonlinearities. 
A phase dynamics may be suitable for the weak nonlinear regime at the mixed ST/ALT pattern (this work is in progress). Moreover, a phase description can be extended to the beating regime (results on nonlocal coupling for this regime will be reported elsewhere~\cite{Miranda208}), although its strongly nonlinear character, except for the defect mediated turbulence displayed in the ST/DW regime. Regarding the Kuramoto model~\cite{Kuramoto03} for a weakly nonlinear system of interacting i-oscillators (for $i=1,\dots,N$) with their respective raw i-phases, $\phi_{i}$, which include the contributions from the fundamental and the harmonic modes, the most suitable model might approach the following phase equation for a local (nonlocal) coupling:

\begin{eqnarray}
\label{eq:phase1} \dot{\phi_{i}}=F_{i}(\omega_i)+\sum_{j=1}^N H_{ij}(t)\cdot \Gamma_{ij}(\phi_i-\phi_j)
\end{eqnarray}

Where $F_i$ is a function of the fundamental frequencies of the $i$-oscillators: $F_i(\omega_v$) in the mixed ST/ALT regime, and $F_i(\omega_v,\omega_z)$ in the beating regimes. $H_{ij}(t)$ accounts for the time-dependent adjacency matrix whose $ij$-elements are nonzero, $H_{ij}(t)=1$, for each connected pair of $ij$-oscillators, while $H_{ij}(t)=0$ otherwise. In our system, this connection defines the topology of the network which happens to be linked to the dynamics (i.e. beats~\cite{Miranda208}, collapsing and spreading phenomena). In fact, $H_{ij}(t)$ is providing the range of interaction, meanwhile $\Gamma_{ij}$ represents the magnitude of the coupling strength between connected oscillators. Near threshold, $\Gamma_{ij}$ usually takes the easiest description~\cite{Kuramoto75}: $\Gamma_{ij}\simeq \sin(\phi_i-\phi_j)$, but in order to follow the dynamics further, a more complex $\Gamma_{ij}$ function might be numerically tested, holding the effect of nonlinearities which are responsible for a stronger coupling strength.\par

Concerning coherent phenomena from an experimental point of view in extended systems, we should emphasize the dynamical features which are also shared with other systems in the route to weak turbulence: (i) the existence of critical domains with saturated widths (stationary clusters in the ST/ZZ regime of size $L_c$), (ii) the existence of a cascade of bifurcations towards a regime (ST/ZZ) where the fundamental traveling modes split in both space and time, and (iii) the existence of localized patterns which are strongly nonlinear (ST/DW). We expect that this results may provide a new insight into spatiotemporal chaos and complex systems. \par

\begin{acknowledgments} 
M.A. Miranda is grateful to P. Collet and R. Ribotta for fruitful comments. This work has been partly supported by the Spanish Contract No. BFM2002-02011 and No. FIS2008-01126, and by PIUNA (University of Navarra, Spain). M.A. Miranda acknowledges financial support from the ``Asociaci\'on de amigos de la Universidad de Navarra''.
 \end{acknowledgments}

\bibliography{bib150109}

\end{document}